\documentclass[runningheads]{llncs}
\usepackage{makeidx}  
\usepackage{amsmath}
\usepackage{amssymb}
\usepackage{xcolor}
\usepackage[colorlinks=true,urlbordercolor=green, linkcolor=red, citecolor=red, filecolor=magenta,menucolor=red, pagecolor=red, urlcolor=black]{hyperref}

\usepackage[letterpaper,hmargin=1.55in,vmargin=1.15in]{geometry}

\begin{document}
\frontmatter          
%
%
%
\mainmatter              
\title{Practical Identity-Based Encryption (IBE) in Multiple PKG Environments and Its Applications}
\titlerunning{Practical IBE in Multi-PKG Environments and Its Applications}
\author{Shengbao Wang}
\institute{Department of Computer Science and Engineering, \\
Shanghai Jiao Tong University \\
800 Dongchuan Road, Shanghai 200240, China\\
\email{shengbao-wang@cs.sjtu.edu.cn}\\
\bigskip
March 20, 2007 }

\maketitle              

\begin{abstract}
In this paper, we present a new identity-based encryption (IBE)
scheme using bilinear pairings. Our IBE scheme enjoys the same
\textsf{Key Extraction} and \textsf{Decryption} algorithms with the
famous IBE scheme of Boneh and Franklin (BF-IBE for short), while
differs from the latter in that it has modified \textsf{Setup} and
\textsf{Encryption} algorithms.

Compared with BF-IBE, we show that ours are more practical in a
multiple private key generator (PKG) environment, mainly due to that
the session secret $g_{ID}$ could be pre-computed \emph{before} any
interaction, and the sender could encrypt a message using $g_{ID}$
prior to negotiating with the intended recipient(s). As an
application of our IBE scheme, we also derive an escrowed ElGamal
scheme which
possesses certain good properties in practice.\\

\textbf{Keywords:} identity-based encryption (IBE), public key
encryption (PKE), escrowed ElGamal, bilinear pairings
\end{abstract}

\section{Introduction}
\label{Introduction}

The idea of \textit{identity(ID)-based cryptography} was first
introduced by Shamir in 1984 \cite{S84}. The basic idea behind an
ID-based cryptosystem is that end users can choose an arbitrary
string, for example their email addresses or other online
identifiers, as their public key. The corresponding private keys are
created by binding the identity with a master secret of a trusted
authority (called private key generation, or PKG for short). This
eliminates much of the overhead associated with key management.

In 2001, Boneh and Franklin \cite{BF01} gave the first fully
functional solution for ID-based encryption (IBE) using the bilinear
pairing over elliptic curves. Based on pairings, Sakai and Kasahara
presented another IBE (SK-IBE for short) scheme by using another
\textsf{Key Extraction} algorithm in 2003 \cite{SK03}. However, the
Boneh-Franklin scheme (BF-IBE for short) has received much more
attention in recent years.

In this paper, we give a new IBE scheme based on bilinear pairings.
Our scheme has the same \textsf{Key Extraction} and
\textsf{Decryption} algorithms with BF-IBE, while differs from the
latter in that it has different \textsf{Setup} and
\textsf{Encryption} algorithms. We show that ours are more practical
in a multiple private key generator (PKG) environment. Parallel to
\cite{BF01}, we also derive an escrowed ElGamal \cite{ElGamal}
encryption scheme from our IBE scheme. Furthermore, we show how the
derived ElGamal encryption enables a dual decrptor public key
encryption (PKE) scheme.

We note that SK-IBE due to Sakai and Kasahara \cite{SK03} has a
better performance than BF-IBE and ours. Especially, SK-IBE are also
very practical in multiple PKG environments. However, its
applicability to some circumstance are not comparable to BF-IBE,
e.g. ,
it seems very hard to derive from it an escrowed ElGamal encryption scheme. In this regard,
we do not compare the new IBE with SK-IBE for now.\\

\noindent \textbf{Paper Organization.} The rest of this paper is
structured as follows. In the next section, we give the necessary
definition for bilinear pairings. Section 3 describes our IBE
scheme. In Section 4, we present a new escrowed ElGamal encryption
scheme. Section 5 contains a brief conclusion and indicates our
ongoing work.

\section{Bilinear Pairings}
\label{Pairings}
In this section, we describe in a more general format the basic
definition and properties of the pairing: more details can be found
in \cite{BF01}.

Let ${\mathbb{G}}_1 $ be a cyclic additive group generated by an
element $P$, whose order is a prime $p$, and ${\mathbb{G}}_2$ be a
cyclic multiplicative group of the same prime order $p$. We assume
that the discrete logarithm problem (DLP) in both ${\mathbb{G}}_1 $
and ${\mathbb{G}}_2 $ are hard.

\begin{definition}
An \textit{admissible pairing} $e$ is a bilinear map
$e:{\mathbb{G}}_1 \times {\mathbb{G}}_1 \to {\mathbb{G}}_2 $, which
satisfies the following three properties:
\begin{enumerate}
    \item \emph{Bilinear}: If $P, Q \in {\mathbb{G}}_1 $ and $a,b \in \mathbb{Z}_p^\ast $, then $e(aP, bQ) =
e(P, Q)^{ab}$;
    \item \emph{Non-degenerate}: $e(P, P) \ne
1$;
    \item \emph{Computable}: If $P, Q \in {\mathbb{G}}_1$, one can compute $e(P, Q) \in {\mathbb{G}}_2 $ in
polynomial time.
\end{enumerate}
\end{definition}

\section{New IBE Scheme and Its Fitness for Multiple PKG Environments}
\label{Gentry's}

For the problem of inherent key escrow, the difficulty of
establishing secure channels for private key distribution, and to
avoid the single point of failure of using only one PKG, it is
well-known that (single-PKG) IBE is only well suitable for use in
relatively small and close organizations, i.e. with each
organization has its own private key generator, generating private
keys for the principal within its domain.

For an IBE to be used in a multiple PKG environment (or, cross
domains), all that is needed is the availability of \emph{standard}
pairing-friendly curves and a common group generator point $P$. We
note that this is a reasonable requirement. In fact, elliptic
curves, suitable group generator points and other cryptographic
tools have been standardized for non-IBE applications, for example
in the NIST FIPS standards \cite{MB05}. Once these group generator
points and curves have been agreed upon, each PKG can generate its
own random master secret.

\subsection{Description of the Scheme}
Let $\mathbb{G}_1$ and $\mathbb{G}_2$ be groups of prime order $p$,
and let $e:{\mathbb{G}}_1 \times {\mathbb{G}}_1 \to {\mathbb{G}}_2 $
be the bilinear pairing. $P$ is
a generator points of $\mathbb{G}_1$. The IBE system works as follows.\\

\noindent \textbf{\textsf{Setup.}} Given a security parameter $k$,
the PKG does the following:

\begin{enumerate}
  \item Chooses a random $s \in
\mathbb{Z}_p$, calculates $P_{Pub}=s^{-1}P \in \mathbb{G}_1$
\footnote{Note that in BF-IBE, the public key of PKG is $P_{Pub}=sP
\in \mathbb{G}_1$ instead.}.
  \item Picks a cryptographic hash functions $H_1 : \{0, 1\}^*\rightarrow
  \mathbb{G}_1^*$, a cryptographic hash
function $H_2: \mathbb{G}_2 \rightarrow \{0, 1\}^n$ for some $n$.
  \end{enumerate}

The message space is $\mathcal{M}= \{0, 1\}^n$. The ciphertext space
is $C = \mathbb{G}_1^*\times \{0, 1\}^n$. The public \emph{params}
are $<q, \mathbb{G}_1, \mathbb{G}_2, e, P, P_{Pub}, n, H_1, H_2>$
and
the \emph{master key} is $s$.\\

\noindent \textbf{\textsf{Key Extraction.}} This algorithm is
identical to that of BF-IBE. To generate a private key for identity
$ID\in \{0, 1\}^*$, the PKG first computes $Q_{ID}= H_1(ID)\in
\mathbb{G}_1^*$, and then sets the private key $d_{ID}$
to be $d_{ID} = sQ_{ID}$ where $s$ is the master key.\\

\noindent \textbf{\textsf{Encryption.}} To encrypt message $m\in
\mathcal{M}$, the sender picks randomly a $r\in \mathbb{Z}_p$, using
the receiver's identity $ID$ to compute $Q_{ID}= H_1(ID)\in
\mathbb{G}_1^*$, sets the ciphertext to be
$$C= \langle rP_{Pub},\  m\oplus H_2(g_{ID}^r)\rangle,\ \text{where}\  g_{ID}=e(P, Q_{ID})\in \mathbb{G}_2^*.$$\\

\noindent \textbf{\textsf{Decryption.}} This algorithm is identical
to that of BF-IBE. To decrypt a ciphertext $C=\langle U, \ V \rangle
\in \mathcal{C}$, using the private key $d_{ID}$ of the identity
$ID$ computes
$$m=V\oplus H_2(e(U,\ d_{ID})).$$

\noindent \emph{Consistence:} The recipient can correctly decrypt
$C$ to get $m$ since
\begin{eqnarray*}
   && e(U,\ d_{ID})\\
  &=&e(rs^{-1}P, \ sQ_{ID}) \\
  &=&e(P, Q_{ID})^r.
\end{eqnarray*}

\subsection{Its Fitness for Multiple PKG Environments}

As mentioned above, an IBE scheme is often used across multiple
PKGs, namely for each organization (e.g., a company), it has its own
PKG. In many cases, a principal may need to encrypt messages to
principals from different domains. For example, for a salesman of
company $A$, he may need to encrypt messages to Bob from company
$B$, Carol from company $C$, or Emmy who he does not know which
company she is belonging to by now.

Now we compare our new IBE with BF-IBE \cite{BF01} in such an
environment. The \textsf{Setup} algorithm in our IBE requires one
more fast inverse operation in $\mathbb{Z}_p$ than BF-IBE, and the
\textsf{Key Extraction} and \textsf{Decryption} algorithms in the
two IBE schemes are the same. In the following, we discuss what
significance our different \textsf{Encryption} algorithm could bring
in practice.

In BF-IBE \cite{BF01}, the session secret, i.e. th term $g_{ID}$ is
computed as $g_{ID} = e(P_{Pub},\ Q_{ID})$, in which $P_{Pub}$ is
the public key of the intended receiver's PKG. We emphasize that in
a multiple PKG environment, before computing the second part of the
ciphertext, i.e. $V$, and especially, the term $g_{ID}$ (requires a
relatively expensive pairing evaluation) which are the main
operations of the overall encryption, BF-IBE requires the sender to
first get to know the following two things:

\begin{itemize}
  \item which organization the receiver is from, \emph{and}
  \item the public key associated with the corresponding PKG.
\end{itemize}
  Compared with BF-IBE, the biggest difference of
our IBE is that in the \textsf{Encryption} algorithm, the terms $V$
and especially, $g_{ID} = e(P,\ Q_{ID})$ are computed independently
from \emph{any} PKG's public key. Consequently, in our IBE, the
sender can compute the pairing (and $V$) \emph{before} getting the
public key of the receiver's PKG, in the case that (s)he knows which
organization the receiver is from. Interestingly, the sender can
even pre-compute $g_{ID}$ and $V$ \emph{before} (s)he knows which
organization the receiver is from!

Therefore, our scheme enables a type of efficient ``on the move" IBE
in a multiple PKG environment, which requires very small on-online
work for the sender (i.e. encryptor).

We emphasize that this feature is particularly useful in (ID-based)
broadcasting (or \emph{multiple-recipient}) encryption scenario,
namely with most of the expensive computation pre-computed, the
overall performance will be upgraded to a large extent.

\section{Escrowed ElGamal Encryption}
Parallel to \cite{BF01}, in this section we introduce a new ElGamal
encryption system in which a single escrow key enables the
decryption of ciphertexts encrypted under any public key.\\

\noindent \textbf{Description of the Scheme:}

Our ElGamal escrow encryption works as follows:

\begin{description}
  \item[Setup.] Given a security parameter $k$, the \emph{escrow authority} (EA) does the following:
\begin{enumerate}
  \item Chooses a random $s \in
\mathbb{Z}_p$, calculates two points $Q_1=sP$ and $Q_2=s^{-1}P \in
\mathbb{G}_1$ \footnote{Note that in BF-IBE, the public key of EA is
one point $Q =sP \in \mathbb{G}_1$ instead.}.
  \item Chososes a cryptographic hash functions $H: \mathbb{G}_2 \rightarrow \{0, 1\}^n$ for some $n$.
\end{enumerate}
The message space is $\mathcal{M}= \{0, 1\}^n$. The ciphertext space
is $C = \mathbb{G}_1^*\times \{0, 1\}^n$. The public \emph{params}
are $<q, \mathbb{G}_1, \mathbb{G}_2, e, n, P, Q_1, Q_2, H>$ and the
\emph{escrow key} is $s$.

\item[Key Generation.] Same as in \cite{BF01}, a user generates a public/private key
pair for herself by picking a random $x\in \mathbb{Z}_q$ and
computing $P_{Pub} = xP\in \mathbb{G}_1$. Her private key is $x$,
her public key is $P_{Pub}$.

\item[Encryption.] To encrypt message $m\in
\mathcal{M}$, the sender picks randomly a $r\in \mathbb{Z}_p$, sets
the ciphertext to be
$$C= \langle rQ_2,\  m\oplus H_2(g^r)\rangle,\ \text{where}\  g = e(P, P_{Pub})\in \mathbb{G}_2^*.$$

\item[Decryption.] To decrypt a ciphertext $C=\langle U, \ V \rangle
\in \mathcal{C}$, using the private key $x$ of the identity $ID$
computes
$$m=V \oplus H_2(e(U,\ xQ_1)).$$

\item[Escrow Decryption.] To decrypt a ciphertext $C=\langle U, \ V \rangle$,
using the escrow key $s$ of the EA computes
$$m=V \oplus H_2(e(U,\ P_{Pub})^s).$$
\end{description}

\noindent \emph{Consistence:} The two recipients can correctly
decrypt $C$ to get $m$ since
\begin{eqnarray*}
   && e(U,\ xQ_1)\\
   &=&e(rQ_2,\ xQ_1)\\
   &=&e(rs^{-1}P,\ xsP)\\
   &=&e(rP,\ xP)\\
    &=&e(P,\ P_{Pub})^r\\
    &=&g^r
\end{eqnarray*}
and
\begin{eqnarray*}
   && e(U,\ P_{Pub})^s\\
   &=&e(rs^{-1}P,\ P_{Pub})^s\\
   &=&e(rP,\ P_{Pub})\\
    &=&e(P,\ P_{Pub})^r\\
    &=&g^r.
\end{eqnarray*}

Compared with the scheme in \cite{BF01}, our escrow ElGamal requires
the EA to publish one more point as its public key. An advantage of
our scheme is that the sender can choose a designated EA (from
multiple EAs) after (s)he finished most of the operations of
encrypting a message. This provides the sender with more flexibility
in practice.\\

\noindent \textbf{A Simple and Direct Application:}

If we look the escrow authority (EA) in the above escrowed ElGamal
scheme as an ordinary principal (who has his/her own private and
public key pair), it can be then used as a \emph{dual decryptor PKE
scheme}, i.e., a single ciphertext can be decrypted
\emph{independently} by two different principals. However, unlike in
conventional setting, we require at least one of the recipient to
publish two points (e.g. $Y_1,\ Y_2$) as his/her public key, in the
form of $Y_1=\alpha P$ and $Y_2 = \alpha^{-1}P$ (assuming $\alpha$
is the private key of the recipient).

A good property of this scheme is that the sender can encrypt the
message before (s)he picks up the second recipient. In other words,
after the encryption has been down, the sender can change his/her
mind on who the second recipient will be.

More interestingly, the sender can efficiently add more such
``second recipient", each time (s)he adds one, only one scalar
multiplication is needed, without any expensive pairing computation.
However, we note that the size of the ciphertext will grow linearly.

\section{Conclusion and Ongoing Work}
\label{Conclusion}

The rapid world-wide development of electronic transactions, largely
associated with the growth of the Internet, stimulates a strong
demand for fast, secure and cheap public key schemes. In this paper,
we gave a practical IBE scheme suitable for multiple PKG
environments. Additionally, we proposed a related escrow ElGamal
encryption scheme.

Ongoing work includes studying the formal security of the proposed
two encryption schemes, namely to prove the security of them in the
random oracle model \cite{BR93-ROM} (provided that the Bilinear
Diffie-Hellman (BDH) problem is hard), and exploring its merits in
constructing Certificate-Based Encryption (CBE) \cite{Gen03} and
Certificateless Public Key Encryption (CL-PKE) schemes \cite{AP03}.
\section*{Acknowledgment}
The author would like to thank Xiaohui Liang and Peng Zeng for many
constructive discussions.


\end{document}